\documentclass[letterpaper, aps,prd,twocolumn,showpacs,floatfix,preprintnumbers,amsfont,amsmath,amssymb,nofootinbib, superscriptaddress]{revtex4}

\usepackage{color}
\usepackage{hyperref}
\usepackage[normalem]{ulem}
\usepackage{url}
\usepackage{float}
\usepackage{ctable}
\usepackage{graphicx}
\usepackage{comment}
\usepackage{mathtools}
\usepackage[alsoload=parsec]{siunitx}
\usepackage{physics}
\usepackage{bm}

\newcommand{\tabline}{\specialrule{.1em}{.05em}{.05em}}

\newcommand{\sk}[1]{}

\def\rmd{\mathrm{d}}

\def\Msun{{\rm M}_\odot}
\def\Mchirp{\mathcal{M}}

\newcommand{\be}{\begin{equation}}
\newcommand{\ee}{\end{equation}}
\newcommand{\ba}{\begin{eqnarray}}
\newcommand{\ea}{\end{eqnarray}}

\newcommand\changed[1]{%
  \bgroup
  #1%
  \egroup
}

\allowdisplaybreaks

\begin{document}

\title{Template Bank for Compact Binary Coalescence Searches in Gravitational Wave Data:\\A General Geometric Placement Algorithm}

\author{Javier Roulet}
\email{jroulet@princeton.edu}
\affiliation{\mbox{Department of Physics, Princeton University, Princeton, NJ, 08540, USA}}
\author{Liang Dai}
\affiliation{\mbox{School of Natural Sciences, Institute for Advanced Study, 1 Einstein Drive, Princeton, NJ 08540, USA}}
\author{Tejaswi Venumadhav}
\affiliation{\mbox{School of Natural Sciences, Institute for Advanced Study, 1 Einstein Drive, Princeton, NJ 08540, USA}}
\author{Barak Zackay}
\affiliation{\mbox{School of Natural Sciences, Institute for Advanced Study, 1 Einstein Drive, Princeton, NJ 08540, USA}}
\author{Matias Zaldarriaga}
\affiliation{\mbox{School of Natural Sciences, Institute for Advanced Study, 1 Einstein Drive, Princeton, NJ 08540, USA}}

\date{\today}


\begin{abstract}
We introduce an algorithm for placing template waveforms for the search of compact binary mergers in gravitational wave interferometer data.
We exploit the smooth dependence of the amplitude and unwrapped phase of the frequency-domain waveform on the parameters of the binary.
We group waveforms with similar amplitude profiles and perform a singular value decomposition of the phase profiles to obtain an orthonormal basis for the phase functions. The leading basis functions span a lower-dimensional linear space in which the unwrapped phase of any physical waveform is well approximated. The optimal template placement is given by a regular grid in the space of linear coefficients.
The algorithm is applicable to any frequency-domain waveform model and detector sensitivity curve. It is computationally efficient and requires little tuning.
Applying this method, we construct a set of template banks suitable for the search of aligned-spin binary neutron star, neutron-star--black-hole and binary black hole mergers in LIGO--Virgo data.
\end{abstract}


\maketitle


\section{Introduction}
\label{sec:intro}

The optimal algorithm to search for known signals in the presence of Gaussian noise is matched-filtering, in which a signal template is cross-correlated with the data and triggers are recorded whenever the correlation exceeds some threshold. In the context of gravitational wave detection \changed{with the LIGO~\cite{LIGOdetector} and Virgo~\cite{Virgodetector} interferometers}, compact binary coalescences are a good example of predictable signals for which we have accurate models, and thus are well suited for matched filtering~\changed{\cite{Dhurandar1994, Allen2012}. 
Indeed, the LIGO and Virgo Collaborations have reported gravitational wave signals from 10 binary black hole (BBH) and one binary neutron star (BNS) mergers during their first and second observing runs~\cite{GWTC-1, GW150914, GW151226, O1catalog, GW170104, GW170608, GW170814, GW170817}, all of which were found by search pipelines based on matched-filtering~\cite{gstlal, PYCBCPipeline} (seven of these BBHs were also found by an unmodeled search~\cite{cWB, GWTC-1}). Searches in the public LIGO--Virgo data by independent groups have found seven additional BBHs~\cite{GW151216, O2_BBH}, additional BBH candidates~\cite{Nitz1OGC} and a BNS candidate~\cite{Nitz2019}, also employing matched filtering.}

Since the source parameters describing the waveform are not known\textit{ a priori}, one needs a bank of waveform templates that adequately cover the parameter space.
The notion of good coverage is characterized by the warranty that any physical waveform the search aims to detect has a sufficiently large match with at least one waveform in the bank.
For example, the LIGO and Virgo Collaborations have aimed at a minimum match of 97\% for any aligned-spin binary merger with component masses between 1 and $\sim 200 \,\rm M_\odot$, for which they require a template bank consisting of $\sim \num{4e5}$ waveforms \cite{Canton2017}. Due to the large number of templates involved, matched filtering is a sizeable computational task.
This means that an efficient bank should not over-cover the parameter space. In other words, the templates should be uniformly spaced with respect to a distance defined in terms of the matched-filtering mismatch between templates (defined in \S\ref{sec:metric}). This incorporates the notion that, from the perspective of signal detection, two waveforms that sufficiently resemble each other are essentially indistinguishable in the presence of noise.
Source parameters can be mutually degenerate in the sense that different parameter combinations may describe similar waveforms.
The optimal placement of templates in physical parameter space is very non-uniform; for example, an order of magnitude more templates are needed to search for mergers with $1$--$3\,\Msun$ components (``neutron stars'') than for mergers with $3$--$200\,\Msun$ (``black holes'').

Two broad classes of template-placement algorithms have been developed in the literature. One robust method is ``stochastic placement''~\citep{Harry:2009ea, Ajith2014, Privitera2014, Capano2016}: waveforms are randomly drawn from the desired parameter space, and one gradually builds up the bank by only accepting newly drawn waveforms that differ sufficiently from the ones the bank already has, and rejecting those that are too similar to at least one existing waveform. Stochastic placement, however, has the shortcoming that a large number of trial waveforms needs to be drawn before convergence is achieved (much more than the required number of templates in the bank). This method also tends to over-cover the parameter space, in the sense that the average template density is higher than optimal at fixed minimum match~\citep{Roy2017a}.

A different method to construct the bank is ``geometric placement''. Here, a metric in the parameter space is defined based on the matched-filtering overlap between waveforms~\citep{Owen1996, Owen1999}.
This metric is then used to define a regular lattice~\citep{Babak2006, Cokelaer2007, Babak2013}. However, it is in general difficult to derive this metric, especially if the parameter space is high dimensional or if the waveform model is not analytic.
Approximations to the metric have first been found by using suitably reparameterized analytic, post-Newtonian (PN) non-spinning waveform models~\citep{Owen1996, Owen1999, Tanaka2000}; later generalizations include the use of phenomenological waveform models and template parameters~\citep{Ajith2008}, the inclusion of aligned-spin PN models~\citep{Brown2012, Harry2014}, or numerical evaluation from arbitrary waveform models~\citep{Roy2017b}.

In practice, a combination of the two methods is often a better strategy. For example, one can place templates geometrically at low masses and stochastically at high masses \citep{Capano2016,Canton2017}, or one can use many small patches with regularly spaced templates, which are themselves placed stochastically to cover the entire parameter space \citep{Roy2017b}.

In this work, we develop a fast and general method to construct a high-effectualness template bank using geometric placement.
Our method relies on the construction of a flat, linear space of orthonormal phase functions that embeds the space of physical waveforms. The Euclidean distance in this space coincides with the mismatch distance between similar waveforms, making these coordinates naturally suited for geometric placement of templates.
\changed{Besides optimal template placement, having this geometrical notion turns out to be helpful for a number of reasons. It allows to refine the bank locally around triggers at the time of search, reducing the amount of templates in the bank at fixed effective coverage. Moreover, a crucial stage of searches involves signal consistency checks, that assess the probability that the residual between a best-fitting template and a candidate signal is explained by Gaussian noise in order to reject non-Gaussian noise transients~\cite{Allen2005, gstlal, PYCBCPipeline, pipelinepaper, vetopaper}. With the bank described here, these tests can be made orthogonal to the linear space of waveforms, so that they are insensitive to mismatches due to the discreteness of the bank. This allows to make the tests more stringent and improves the sensitivity of the search~\cite{vetopaper}.}
We further require that the template bank be built from sub-banks that can be approximated to have a fixed amplitude profile $A(f)$. This feature is useful for implementing the noise amplitude-spectral-density drift correction, a key component for precise matched filtering \cite{pipelinepaper, psddriftpaper}.
\changed{Together, these analytical properties make our template bank appealing, even considering that there are other template banks with comparable effectualness and number of templates in the literature.
Finally, building a new template bank enables us to customize a number of other choices, like the frequency range and parameter space covered, in the context of our search pipeline~\cite{pipelinepaper} and the detector performances during the observation time analyzed.}
The coordinates presented in this work are similar in essence to the ones introduced in \citet{Brown2012}, except that we generalize them to arbitrary waveform models and component mass ranges.

The paper is organized as follows. In \S\ref{sec:metric} we define a metric based on the mismatch between templates and show how the desired Euclidean space can be constructed. In \S\ref{sec:tb} we apply this formalism to the construction of a search-quality template bank that targets stellar-mass compact binary mergers. We summarize our results in \S\ref{sec:conclusions}.
The bank presented here was used in the searches described in Refs.~\cite{pipelinepaper, O2_BBH}, except that the bank used in \cite{pipelinepaper} had some limited differences that we report in Appendix~\ref{app:tb_differences}.

\section{Linear metric space} 
\label{sec:metric}

In this section, we define the notion of distance between templates and describe the construction of a low-dimensional linear space of phase functions in which the metric is Euclidean. We build this linear space based on the intuition that the unwrapped phases are smooth functions of the wave frequency\changed{~\cite{Cutler1994}} and hence are linear combinations of a small number of basis functions~\changed{\cite{Tanaka2000, Brown2012}}.

\subsection{Mismatch distance}

We first introduce the noise-weighted inner product in the frequency domain \citep{Allen2012}
\begin{equation} \label{eq:ip}
    (h_i \mid h_j) \vcentcolon= 4 \int_0^\infty \rmd f\, \frac{\tilde h_i(f) \tilde h_j^\ast(f)}{S_n(f)}.
\end{equation}
Here, $S_n(f)$ is a fiducial one-sided power spectral density (PSD) of the detector noise and tildes indicate Fourier transforms. 
The match between $h_i$ and $h_j$ is given by $\Re z_{ij}$, where
\begin{equation} \label{eq:z_ij}
    \begin{split}
        z_{ij} &\vcentcolon= \frac{(h_i \mid h_j)}{\sqrt{(h_i \mid h_i) (h_j \mid h_j)}} \\
        &\equiv (h_i \mid h_j).
    \end{split}
\end{equation}
In the second line, we normalize the waveforms to
\begin{equation} \label{eq:norm}
    (h \mid h) = 1,
\end{equation}
as usually the template waveforms are defined up to an overall normalization.
Since all possible coalescence times and phases are searched for, waveforms related by time and phase offsets are described by the same waveform in template bank. Thus, the match is maximized over time and phase offsets:
\begin{equation} \label{eq:match}
    \begin{split}
        m_{ij} &\vcentcolon= \max_{\tau_0, \phi_0} \big(\Re z_{ij}(\tau_0, \phi_0)\big) \\
        &= \max_{\tau_0} \abs{z_{ij}(\tau_0)},
    \end{split}
\end{equation}
where $\tau_0$ and $\phi_0$ are the time and phase offsets between $h_i$ and $h_j$, respectively.
We define the mismatch distance $d_{ij}$ between the two waveforms by
\begin{equation} \label{eq:distance}
    d_{ij}^2 = 1 - m_{ij}.
\end{equation}
We seek a parametrization of waveforms under which the mismatch distance has an Euclidean metric for similar waveforms.

\subsection{Linear space}

\begin{figure}
    \centering
    \includegraphics[width=1\linewidth]{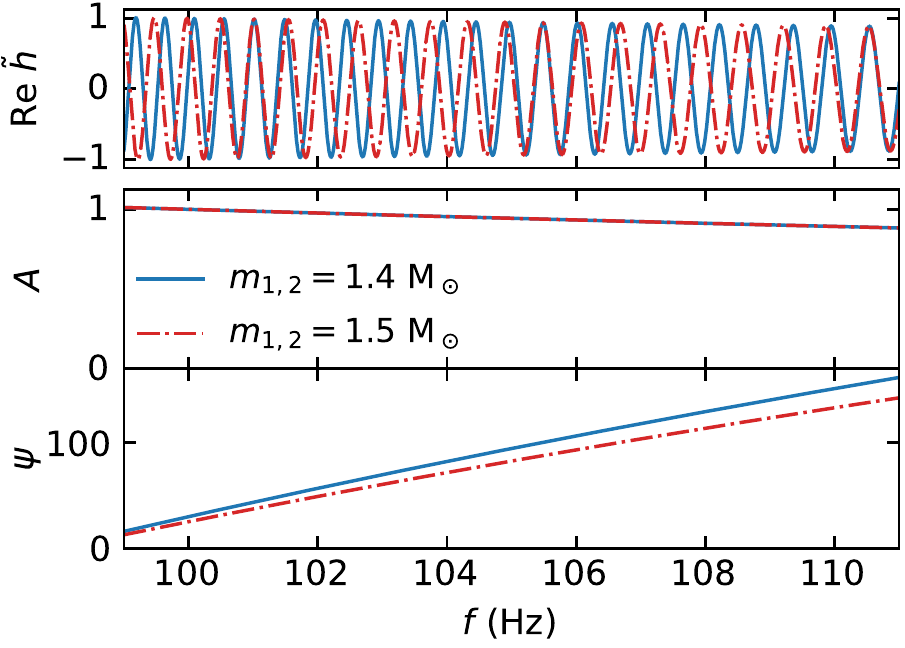}
    \caption{An example of two waveforms that look very different to each other in the frequency domain (top panel) but have very similar amplitude and phase profiles (middle and bottom panels). The amplitude and phase profiles can be well captured by a low-dimensional linear space spanned by a few basis functions. Waveform amplitudes are shown in arbitrary units.}
    \label{fig:amp_phase}
\end{figure}

A general frequency-domain waveform model can be cast to the form
\begin{equation}
    \tilde h(f;\bm{p}) = A(f;\bm{p}) e^{i \psi(f;\bm{p})}.
\end{equation}
\changed{Under the approximation that the dominant mode of gravitational radiation has $(\ell, \abs{m})=(2, 2)$ and that spin-orbital precession and eccentricity effects are insignificant,} the frequency dependent functions $A$ and $\psi$ vary slowly with the binary parameters $\bm{p}$, as illustrated in Fig.~\ref{fig:amp_phase}. For matched filtering, the phase $\psi$ is the most important to describe with high accuracy, since loss of phase coherence leads to a rapid degradation of the signal-to-noise ratio (SNR).

Moreover, it is important to analyze templates with different amplitude profiles $A(f)$ separately as the matched filtering correction from PSD drifts depends on $A(f)$~\cite{psddriftpaper}. Thus we assume in the following that $A(f;\bm{p})\approx \overline A(f)$ is valid for a suitably chosen subset of parameters. To achieve this, we sort a large number of randomly sampled physical input waveforms into groups of similar amplitude profile.
In each group, we require that the match of the amplitudes to a reference $\overline A(f)$ exceeds a minimum
\begin{equation} \label{eq:amp_match}
    (A_i \mid \overline A) = 4\int_0^\infty \rmd f \frac{A_i(f) \overline A(f)}{S_n(f)} \geqslant 0.96
\end{equation}
for all input waveforms $h_i$ in the group. Note that the match of the amplitudes sets an upper bound on the match of the waveforms. 
Our approach will be to split a template bank into ``subbanks'', each subbank describing one group of input waveforms which share the same approximate amplitude profile $\overline A(f)$.

We design the subbanks in order to minimize the average amplitude mismatch as follows. We start with a single subbank that contains all the input waveforms, and define its reference amplitude profile as the root-mean-square
\begin{equation} \label{eq:amp_rms}
    \overline A(f) \coloneqq \sqrt{\langle A^2 \rangle(f)},
\end{equation}
where the angled brackets indicate average over the input waveforms in the subbank. 
This choice inherits the normalization of the input waveforms.
We compute the amplitude match Eq.~\eqref{eq:amp_match} for all the waveforms; if the worst match satisfies the chosen bound we stop. If it does not, we add a new subbank with a reference amplitude given by the waveform with the worst amplitude match.
We then optimize the choice of reference amplitudes using the $k$-means algorithm: we reassign waveforms to subbanks by their best amplitude match, redefine the amplitude profile of the subbanks using Eq.~\eqref{eq:amp_rms}, and iterate these two steps a few times to achieve convergence. Finally we recompute the worst match and decide if a new subbank is needed, in which case we repeat the above process.

Having decided on the division of subbanks, we wish to find an efficient representation of the set of phases $\psi(f)$ as a linear combination of a small number of basis functions, 
\begin{equation} \label{eq:linear_combination}
    \psi(f;\bm{p}) = \overline\psi(f) + \sum_{\alpha}^\text{few} c_\alpha(\bm{p}) \psi_\alpha(f),
\end{equation}
where $\alpha$ is an integer index that enumerates the basis functions and $\overline\psi(f)$ is an average phase which we are free to define. From now on, we abandon the physical parameters $\bm{p}$ and describe the waveforms in terms of their $c_\alpha$ components:
\begin{equation} \label{eq:calpha_decomposition}
    h(f; \bm{c}) = \overline A(f) \exp\Big[i \Big(\overline\psi(f) + \sum_\alpha c_\alpha\, \psi_\alpha (f) \Big)\Big].
\end{equation}

We now express the match between two waveforms using the above decomposition. As mentioned earlier, template waveforms are defined up to arbitrary time and phase offsets, namely an additive piece to the phase that is a linear function of the frequency $\Delta \psi(f) = \phi_0 + 2 \pi f \tau_0$. We choose the first two basis functions $\psi_0(f)$ and $\psi_1(f)$ to span the subspace of linear phases so that $c_0$ and $c_1$ capture phase and time offsets, respectively, and in particular $\psi_0(f) \equiv 1$.
If two waveforms are similar, their inner product Eq.~\eqref{eq:ip} to second order in $\delta c_\alpha$ is approximately 
\begin{multline} \label{eq:dot}
    \big( h(\bm{c}) \,\big|\, h(\bm{c} + \bm{\delta c}) \big)
    = 4\, \int_0^\infty\, \rmd f\, \frac{\overline A^2(f)}{S_n(f)}
        e^{i \sum_\alpha \delta c_\alpha \psi_\alpha(f)}\\
    \approx 4\,e^{i \delta c_0}\,\int_0^\infty\, \rmd f\, \frac{\overline A^2(f)}{S_n(f)}
        \bigg[1 + i \sum_{\alpha \geqslant 1} \delta c_\alpha\,\psi_\alpha(f) \\
        -\frac{1}{2}\,\sum_{\alpha, \beta \geqslant 1}\,\delta c_\alpha\,\delta c_\beta\,\psi_\alpha(f)\,\psi_\beta(f) \bigg] + \mathcal O(\delta c^3).
\end{multline}
This motivates a new inner product, with respect to which we will orthonormalize the basis functions:
\begin{equation} \label{eq:phase_ip}
\begin{split}
    \langle\psi_\alpha, \psi_\beta\rangle &\vcentcolon= 4\, \int_0^\infty\, \rmd f\, \frac{\overline A^2(f)}{S_n(f)}
        \,\psi_\alpha(f)\, \psi_\beta(f) \\
        &\equiv \delta_{\alpha\beta},
\end{split}
\end{equation}
which we enforce by a suitable choice of the basis functions $\psi_\alpha(f)$ (described below). In particular, the first condition $\langle \psi_0, \psi_0 \rangle = 1$ is the normalization Eq.~\eqref{eq:norm}, and the two first basis functions are
\begin{equation} \label{eq:psi_01}
    \begin{split}
        \psi_0(f) &= 1, \\
        \psi_1(f) &= \frac{f - \overline f}{\sqrt{\overline{f^2} - \overline f^2}},
    \end{split}
\end{equation}
where we define $\overline{f^n} \coloneqq 4\int_0^\infty \rmd f f^n A^2(f)/S_n(f)$.

Using orthonormality, Eq.~\eqref{eq:dot} becomes
\begin{equation}
    \big( h(\bm{c}) \,\big|\, h(\bm{c} + \bm{\delta c}) \big)
      \approx e^{i \delta c_0} \Big(1 - \frac 12 \sum_{\alpha \geqslant 1}\delta c_\alpha^2 \Big)
      + \mathcal O(\delta c^3).
\end{equation}
Thus, for nearby templates the distance Eq.~\eqref{eq:distance} is
\begin{equation} \label{eq:Euclidean}
    d^2_{\bm{c}, \bm{c} + \bm{\delta c}} \approx \frac 12 \sum_{\alpha \geqslant 2}\delta c_\alpha^2 + \mathcal O(\delta c^3),
\end{equation}
which means that the mismatch distance is given by an Euclidean metric in $\bm c$ space at small displacements.
We construct the bank on a regular grid in $\bm c$ space with spacings $\Delta c_\alpha \lesssim 1$, chosen sufficiently small so as to guarantee a minimal loss of match.

We note in passing that we can also compute the distance in the opposite limit of large separation, which is useful for estimating the long-range correlations between triggers from different templates during a search. Assuming now that the templates are separated by $\bm{\delta c} = D \bm{\hat{n}}$, with $\sum_\alpha n_\alpha^2 = 1$ and $D \gg 1$, we can perform a stationary phase approximation around the frequencies $f_j$ at which $\sum_\alpha n_\alpha \psi_\alpha'(f_j)=0$. This yields
\begin{multline}
    \big(h(\bm{c}) \,\big|\, h(\bm{c} + D \bm{\hat n}) \big) =
        4 \int_0^\infty \rmd f \frac{\overline A^2(f)}{S_n(f)} e^{i D \sum_\alpha n_\alpha \psi_\alpha(f)} \\
    \approx \frac{4}{\sqrt D} \sum_j \frac{\overline A^2(f_j)}{S_n(f_j)} \frac{\sqrt{2 \pi} \exp(-i \frac{\pi}{4} + i D \sum_\alpha n_\alpha \psi_\alpha(f_j))}
    {\sqrt{\sum_\alpha n_\alpha \psi_\alpha''(f_j)}}.
\end{multline}
Thus, the long-range correlation between two templates separated by $D$ decays as $1 / \sqrt D$ (this holds for the match without maximization over time).

\begin{figure}
    \centering
    \includegraphics[width=\linewidth]{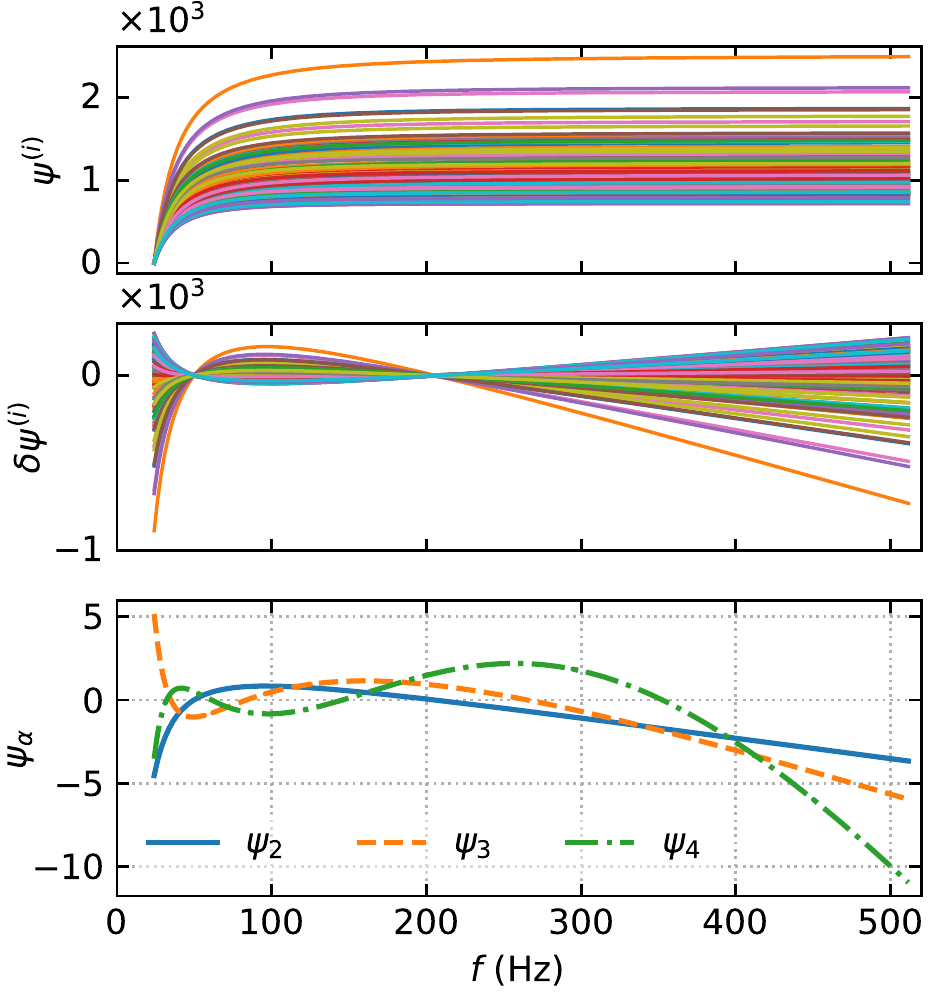}
\caption{Construction of the basis functions $\psi_\alpha$.\textit{ Top panel:} (a subset of 100) input unwrapped phases for random parameters.\textit{ Middle panel:} phase residuals after subtraction of the average phase and orthogonalization with respect to time and phase offsets.
\textit{ Bottom panel:} first three basis functions.}
    \label{fig:psi}
\end{figure}

In practice we choose the set of basis functions $\psi_\alpha(f)$ as follows:
\begin{enumerate}
    \item Define a discrete frequency grid $\{f_k\}$ (our choice is described in \S\ref{sec:tb}). The integrals over frequency will be approximated by quadratures $\sum_k \Delta f_k \ldots$;
    
    \item Compute a moderately large number of waveforms for random parameter choices (we use \num{5e4}), and extract the unwrapped phases, $\{\psi^{(i)}(f_k)\}$, as illustrated in the top panel of Fig.~\ref{fig:psi};
    
    \item Subtract the average phase $\overline\psi(f)$;
    
    \item 
    Subtract the projection onto the first two dimensions so that the phase residuals
    \begin{equation}
        \delta\psi^{(i)}(f) = \psi^{(i)}(f) - \overline\psi(f) 
        - \sum_{\alpha=0}^1 \langle \psi^{(i)} - \overline \psi, \psi_\alpha \rangle\, \psi_\alpha(f)
    \end{equation}
    are orthogonal to $\psi_0$ and $\psi_1(f)$ with respect to the inner product Eq.~\eqref{eq:phase_ip} (middle panel of Fig.~\ref{fig:psi});
    
    \item Construct a matrix of weighted phase residuals 
    \begin{equation}\label{eq:Mik}
      \begin{split}
        M_{ik} &= w_k\, \delta\psi^{(i)}(f_k), \\
        w_k &= 2\, \overline A(f_k)\, \sqrt{\Delta f_k / S_n(f_k)},
      \end{split}
    \end{equation}
    and find its singular-value decomposition (SVD)
    \begin{equation} \label{eq:M}
        M_{ik} = \sum_\alpha U_{i \alpha} D_\alpha V_{\alpha k}.
    \end{equation}
    $U, V$ are orthogonal matrices and we sort the axes so that the eigenvalues $D_\alpha>0$ are in decreasing order. From the orthogonality of $V$, i.e. $\sum_k V_{\alpha k} V_{\beta k} = \delta_{\alpha \beta}$, we can identify
    \begin{equation}
        V_{\alpha k} = w_k\, \psi_\alpha(f_k)
    \end{equation}
    which satisfies the orthonormality Eq.~\eqref{eq:phase_ip} and defines the basis functions, with the convention that the $\alpha$ start at 2 (bottom panel of Fig.~\ref{fig:psi}).
\end{enumerate}

From Eqs.~\eqref{eq:linear_combination} and \eqref{eq:M} it follows that the components of the input waveforms are
\begin{equation} \label{eq:calpha}
    c_\alpha^{(i)} = U_{i \alpha}\, D_\alpha.
\end{equation}
Since $U$ is an orthogonal matrix, $\abs{U_{i\alpha}} \leqslant 1$ and $\abs{c_\alpha} \leqslant D_\alpha$, that is, the extent spanned by the input samples along each dimension in component space is bounded by $D_\alpha$. This means that the information in the templates is captured by the first few components along the larger dimensions, and we can reduce the dimensionality of our description by dropping the dimensions that have $D_\alpha \ll 1$.

\section{Constructing a search quality template bank} 
\label{sec:tb}

In this Section, we apply the method developed in \S\ref{sec:metric} to the construction of a template bank suitable to the search of gravitational wave strain signals from binary neutron stars, neutron-star--black-hole and binary black hole mergers.

We choose lower and upper frequency cutoffs of $f_{\rm min} = \SI{24}{\hertz}$, and $f_{\rm max} = \SI{512}{\hertz}$, respectively. These cutoffs are chosen such that the resulting loss in SNR$^2$ is lower than 2\% for binary neutron star templates (the amplitude profiles of these whitened waveforms, i.e., $A(f)/\sqrt{S_n(f)}$, are essentially independent of parameters since the cut-off scale is outside the LIGO sensitivity band). Formally, the accumulated $\textrm{SNR}^2 = 4\,\int_0^{\SI{24}{\hertz}} \rmd f A^2(f)/S_n(f) \approx 4\int_{\SI{512}{\hertz}}^\infty \rmd f A^2(f)/S_n(f) \approx \num{e-2}$ outside our frequency range.
It is advisable to restrict the frequency range because the linear-free phase, and thus the basis functions, grow rapidly at both ends (see Fig.~\ref{fig:psi}), and our Taylor expansion Eq.~\eqref{eq:dot} would become inaccurate.
As we noted above, it is exactly at these frequencies where the contribution to the matched-filtering SNR vanishes. 
It is better to discard these frequencies rather than to try and capture the negligible information content within by adding extra dimensions to the template bank. 
Furthermore, this has the additional benefit that the strain data can be down-sampled during analysis, which reduces the computational cost of the search.

We define the fiducial PSD empirically from the PSDs of 200 LIGO Handford and LIGO Livingston data files chosen randomly from the Second Advanced LIGO Observing Run (O2) release~\citep{gwosc_url,gwosc}. Each individual PSD was computed as described in \cite{pipelinepaper}. The fiducial PSD is constructed using the 10th percentile of all the sample PSDs in each frequency bin. This choice is robust to large fluctuations in the sample PSDs, and is representative of optimal detector conditions.

We choose a target parameter space of compact binary mergers satisfying the following bounds:
\begin{align}
    & 1\,\Msun < m_1,\,m_2 < 100\,\Msun, \\
    & q > \begin{cases}
               1/50  &\quad \text{if } m_2 < 3\,\Msun \\
               1/18 &\quad \text{otherwise},
           \end{cases} \\
    & \abs{\chi_1}, \abs{\chi_2} < 0.99,
\end{align}
where $m_1$ and $m_2$ are the primary and secondary masses, respectively, $q = m_2 / m_1 \leqslant 1$ is the mass ratio, and $\chi_1$ and $\chi_2$ are the individual dimensionless spin projections in the direction of the orbital angular momentum.
\changed{
The parameter ranges and approximant used are not a constraint from the LIGO and Virgo detectors or the method presented here, but a documentation of the choices we made.
In particular, the mass ratio cut for BBHs was due to the calibration regime of the \texttt{IMRPhenomD} approximant~\cite{Khan2016}.
For NSBH, we extend the maximal $q$ because a substantial part of the NSBH parameter space lies outside the calibrated range of \texttt{IMRPhenomD}. For the purpose of signal detection (as opposed to parameter estimation), the calibration tolerance is less stringent, as long as a signal can be recovered by the model with some combination of parameters.
Compared to other template banks in the literature, the one presented here covers a larger spin range for low-mass objects. Indeed, bounds of $\abs{\chi} < 0.05$~\cite{Canton2017, Brown2012, Roy2017b} or $\abs{\chi} < 0.4$~\cite{Brown2012} have been used in the BNS mass range, the former motivated by the known binary neutron star spins and the latter by the known pulsar spins~\cite{Miller2015}. Neutron stars can in principle have dimensionless spins up to a mass-shedding limit of $\abs{\chi} \sim 0.7$ \cite{Lo2011, Tacik2015}. Other types of compact objects, including light black holes, may in principle have even higher spins. This motivates us to cover this unexplored part of the parameter space.
}

As mentioned before, the number of templates required to describe waveforms from low-mass mergers is significantly larger than that for high-mass mergers, due to the larger number of wave cycles in band. Searches with larger template banks suffer a penalty in sensitivity because of the increased look-elsewhere effect. To prevent the high penalty inherent to the lower-mass region of parameter space from affecting the higher-mass regions, we propose to divide the search space into a number of regions and perform an independent search in each. Each search then only pays an additional look-elsewhere penalty that a few other searches are performed, but is unaffected by the potentially huge size of the other banks. This division can be interpreted as implementing a prior about which templates are more likely to produce an astrophysical trigger: if we expect comparable numbers of high- and low-mass signals but have vastly more templates at low-mass, any particular low-mass template is much less likely to produce an astrophysical trigger.
In addition, templates in different regions of parameter space are sensitive to different types of noise transients in the strain data. Dividing the search into several regions enables us to recognize the different types of noise background that a search using each class of templates is subject to.

\begin{figure}
    \centering
    \includegraphics[width=\linewidth]{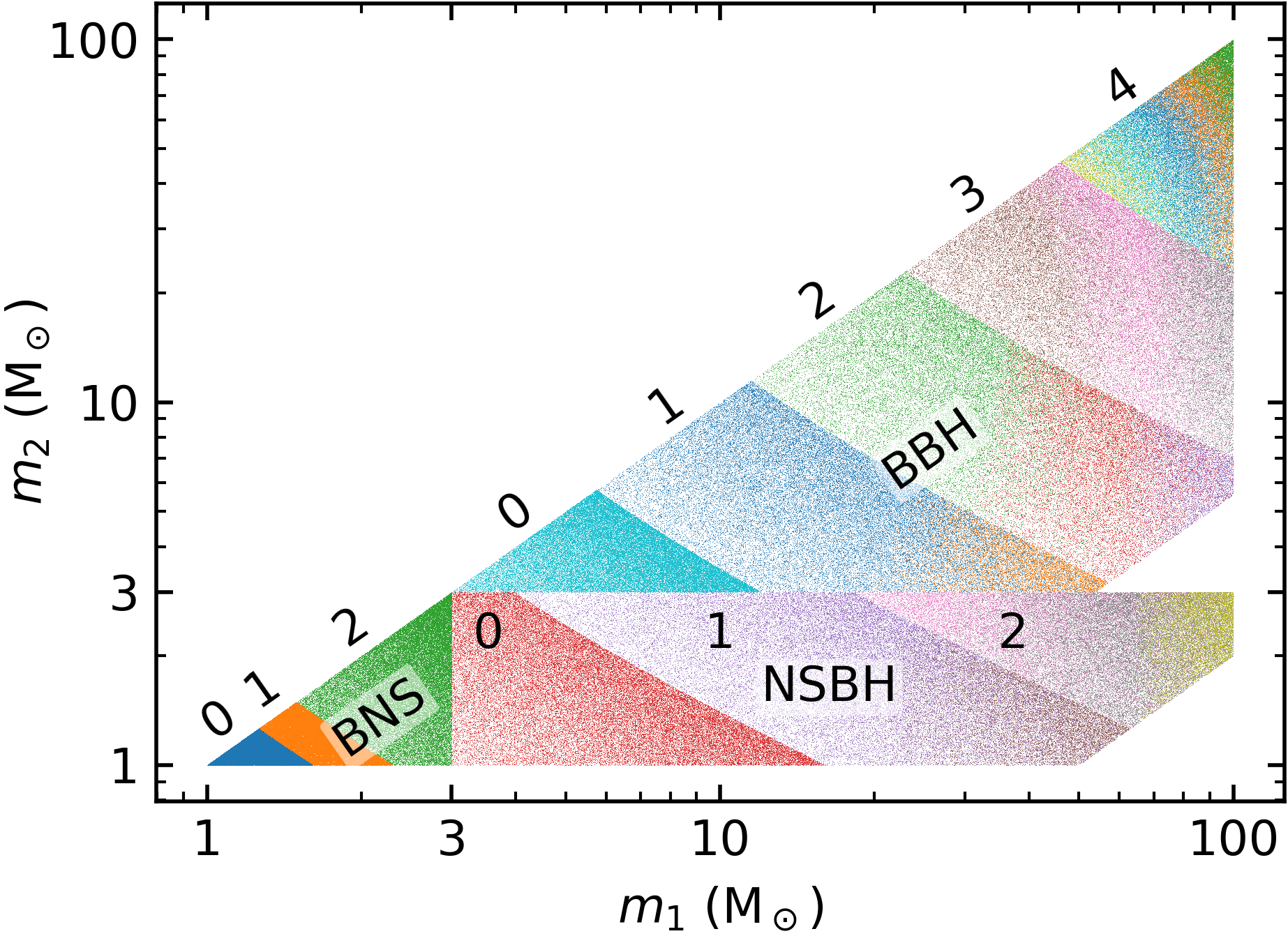}
    \caption{Component masses of the input waveforms used for constructing our template banks. We divide the parameter space according to the component masses into three banks of binary neutron star waveforms (\texttt{BNS 0-2}), three banks of neutron-star--black-hole waveforms (\texttt{NSBH 0-2}), and five banks of binary black hole waveforms (\texttt{BBH 0-4}). We further divide each bank into subbanks (color coded) under the criterion that the match of the amplitude profiles with a reference profile specific to each subbank exceeds $0.96$. The dots show the \num{5e4} input waveforms we use to build each bank. We do not scatter plot here individual template waveforms in each bank, as those lack an explicit representation in terms of the source physical parameters.}
    \label{fig:all_banks}
\end{figure}

Under the above motivations, we divide the search space into regions based on the component masses, and construct a separate template bank for each of them. The division is illustrated in Fig.~\ref{fig:all_banks} and is defined as follows. We refer to binary components with masses between 1 and $3\,\Msun$ as neutron stars, and to components with masses between 3 and $100\,\Msun$ as black holes. We make three binary neutron star template banks, three neutron-star--black-hole (NSBH) banks, and five binary black hole banks. The banks within each of these categories are defined by bins in the chirp mass $\Mchirp \coloneqq (m_1\,m_2)^{3/5}/(m_1 + m_2)^{1/5}$. We put the bounds between the three BNS banks at $\Mchirp = \{1.1, 1.3\}\,\Msun$. This choice is motivated by the observation that the chirp masses of the known Galactic binary neutron stars expected to merge within a Hubble time lie in a narrow range \citep{Farrow2019}, and therefore we might expect more astrophysical signals from this chirp mass range (which we further expand to account for the redshift of the detector-frame masses up to $z \sim 0.05$, or a luminosity distance $d_L \sim \SI{200}{\mega\parsec}$). In this way, we minimize the number of templates in the most astrophysically probable BNS bank, \texttt{BNS 1}, enhancing our sensitivity to those systems\footnote{The authors thank Thomas Dent for suggesting this approach.}. A similar strategy was adopted in Refs.~\citep{Magee2019, Nitz2019}. For other banks, we use logarithmic chirp-mass bins: we place the bounds between the three NSBH banks at $\Mchirp = \{3, 6\}\,\Msun$, and those between the five BBH banks at $\Mchirp = \{5, 10, 20, 40\}\,\Msun$.
We generate \num{5e4} input waveforms in each bank using the \texttt{IMRPhenomD} approximant \citep{Khan2016}. Based on the amplitude profiles $A_i(f)$ of the input waveforms, we further divide each bank into subbanks as explained in \S\ref{sec:metric}.
We find that a single subbank is sufficient for waveforms with $m_{1,2} \lesssim 15\,\Msun$, but multiple amplitude subbanks are needed for heavier mergers as the frequency at which $A(f)$ is cut-off falls within the LIGO sensitive band.  
Table~\ref{tab:banks} summarizes the parameters of all template banks. The banks differ greatly in size, which justifies the division of the search space into multiple banks.

\begin{table*}
    \centering
    \caption{Summary of the parameters of the template banks. Columns 2 to 6 describe the bounds of physical parameter space that each bank is designed to cover. $\zeta$ is a tunable fudge parameter that controls the tolerance for removing nonphysical grid points. $\Delta c_\alpha$ is the grid spacing that we chose for each bank. $N_{\rm subbanks}$ is the resulting number of subbanks in each bank. $d_{\rm subbanks}$ and $L_{\rm max, subbanks}$ are the dimensionalities of each subbank (sorted by increasing mean total mass) and the size of their largest dimension, respectively. $N_{\rm templates}$ is the total number of templates in each bank.}
    \begin{tabular}{l|*{5}{c}|*{2}{S[table-format=1.2]}|cllS[table-format=6.]}
    \tabline
    \tabline
         Bank & $m_1 \,(\Msun)$ & $m_2 \,(\Msun)$ & $\Mchirp \,(\Msun)$ & $q_{\rm min}$ & $\abs{\chi_{1,2}}_{\rm max}$ & {$\zeta$} & {$\Delta c_\alpha$} & $N_{\rm subbanks}$ & $d_{\rm subbanks}$ & $L_{\rm max, subbanks}$ & {$N_{\rm templates}$} \\
         \tabline
         \texttt{BNS 0} & & & ${} < 1.1$ & & & & & 1 & $2$ & $777.0$ & 48806 \\
         \texttt{BNS 1} & $(1, 3)$ & $(1, 3)$ & $(1.1, 1.3)$ & --- & 0.99 & 0.05 & 0.55 & 1 & $2$ & $434.3$ & 23856 \\
         \texttt{BNS 2} & & & ${} > 1.3$ & & & & & 1 & $2$ & $824.6$ & 43781 \\
         \hline
         \texttt{NSBH 0} & & & ${} < 3$ & & & & & 1 & $4$ & $753.4$ & 84641 \\
         \texttt{NSBH 1} & $(3, 100)$ & $(1, 3)$ & $(3, 6)$ & $1/50$ & 0.99 & 0.05 & 0.5 & 2 & $6,6$ & $259.5, 166.8$ & 85149 \\
         \texttt{NSBH 2} & & & ${} > 6$ & & & & & 3 & $5,4,4$ & $87.5, 61.2, 9.4$ & 15628 \\
         \hline
         \texttt{BBH 0} & & & ${} < 5$ & & & & 0.55 & 1 & $3$ & $270.6$ & 8246 \\
         \texttt{BBH 1} & & & $(5, 10)$ & & & & 0.55 & 2 & $4,4$ & $113.7, 50.0$ & 4277 \\
         \texttt{BBH 2} & $(3, 100)$ & $(3, 100)$ & $(10, 20)$ & $1/18$ & 0.99 & 0.05 & 0.5 & 3 & $3,4,3$ & $41.5, 33.5, 10.3$ & 1607 \\
         \texttt{BBH 3} & & & $(20, 40)$ & & & & 0.45 & 3 & $2,2,2$ & $11.7, 10.8, 4.9$ & 225 \\
         \texttt{BBH 4} & & & ${} > 40$ & & & & 0.35 & 5 & $2,2,2,1,1$ & $2.9, 2.0, 1.1, 0.7, 0.5$ & 46 \\
         \tabline
         Total&\multicolumn{10}{l}{} & 316262 \\
         \tabline
         \tabline
    \end{tabular}
    \label{tab:banks}
\end{table*}

\begin{figure}
    \centering
    \includegraphics[width=\linewidth]{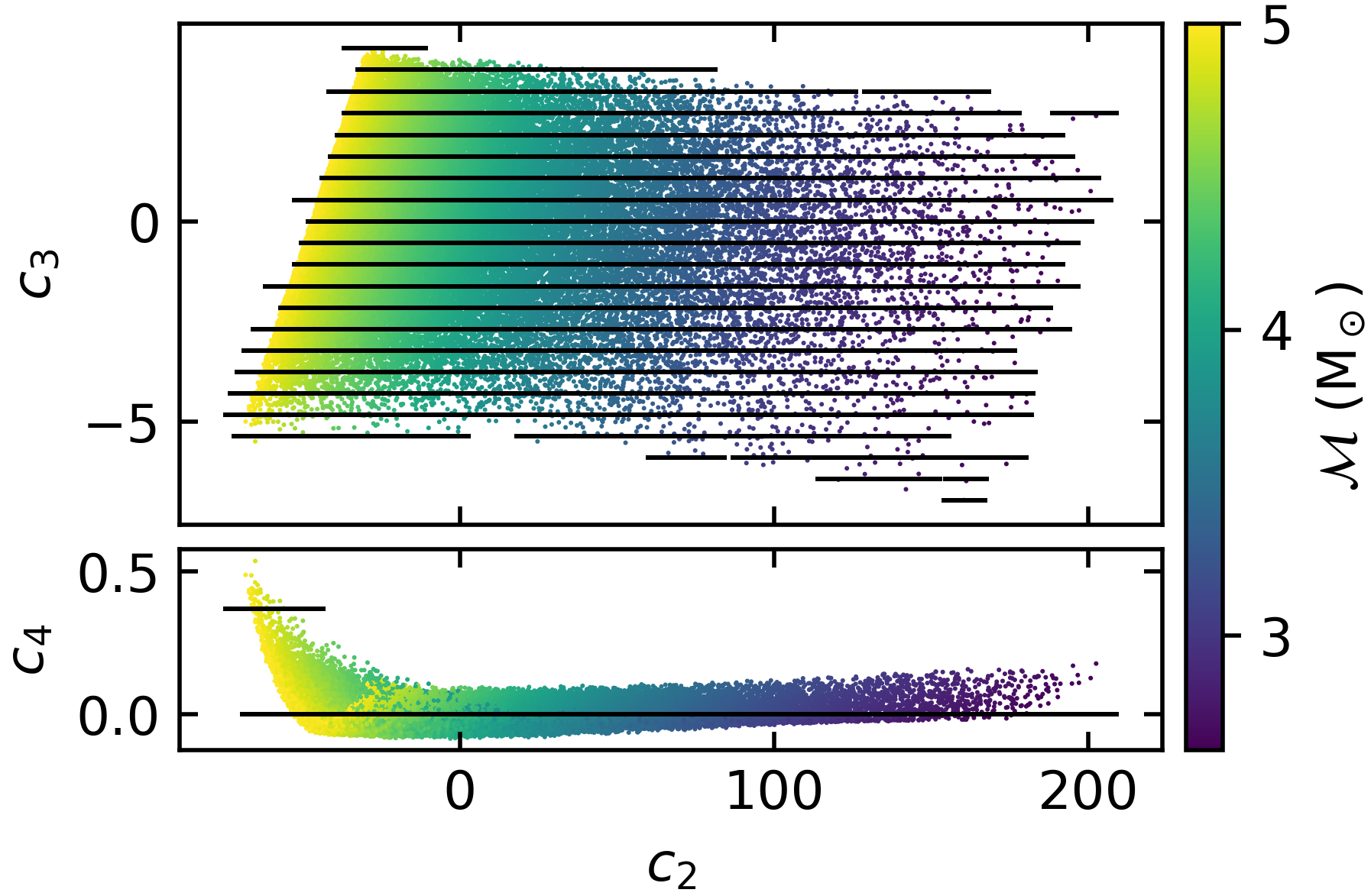}
    \caption{Our three-dimensional \texttt{BBH 0} template bank in the component space (black dots, which at this scale appear as lines), projected along two different axes. Underlaid are the input waveforms used to build the bank, projected according to Eq.~\eqref{eq:calpha} and color-coded by their chirp mass.}
    \label{fig:calpha}
\end{figure}

For each subbank, we apply the procedure outlined in \S\ref{sec:metric} to define a set of basis phase functions that generate a linear space and obtain the projections of the input waveforms onto that space. These are shown in Fig.~\ref{fig:calpha} for the example case of the \texttt{BBH 0} bank, with the points color-coded by their chirp mass. The first three dimensions capture practically all the diversity of the input waveforms. Also note the large differences in size from the leading dimension to the sub-leading ones. \changed{The number of cycles, proportional to $\mathcal M^{-5/3}$, is the best-measured parameter and thus should approximately correspond to the coefficient of the leading dimension~\cite{Cutler1994, Dhurandar1994}. Indeed, this is observed in Fig.~\ref{fig:calpha}, confirming that the decomposition is working as expected.}

Next, we choose a grid spacing $\Delta c_\alpha$ common to all dimensions and define a rectangular grid in component space as follows. We force the point $\bm{c} = \bm{0}$ to be a grid point, because the SVD typically aligns the highest density regions (where the input physical waveforms tend to be) with the axes. Along each dimension, we add uniformly-spaced points until the whole range spanned by the input waveforms is covered. We allow the spacing to slightly decrease so that the most extreme input component is half the grid spacing away from the most extreme grid point. We do this for each dimension and in the positive and negative directions separately. Finally, not all the points of the rectangular grid describe physically viable waveforms. We only keep the templates that are close to at least one input waveform, with the following criterion.
For every input waveform set of components, we keep the closest grid point and a patch of the grid around it, with size equal to the corresponding dimension times a tunable fudge factor $\zeta \sim 0.1$.

Indeed, as Fig.~\ref{fig:calpha} shows, the input physical waveforms do not fill the entire rectangular volume but are distributed within some irregularly shaped region. Furthermore, the density of input waveforms is low in the low-$\Mchirp$ region, where the waveforms have more wave cycles in band and hence are mutually more distinguishable. Holes can be produced in the physically viable region if the fudge factor $\zeta$ is too small, and there is an excess of unphysical templates if $\zeta$ is too large. We choose the $\Delta c_\alpha$ and $\zeta$ parameters such that we achieve a good balance between economic template bank size and high bank effectualness. The values chosen for each bank are reported in Table~\ref{tab:banks}.

In Table~\ref{tab:banks} we observe a general trend with the mass: the banks for lighter mergers tend to have fewer subbanks and the first dimension spans a wider range. By comparison, the banks for heavier mergers have more subbanks, with smaller dimensions. The increase in the number of subbanks for heavier mergers is caused by the cutoff frequency falling in the band, which increases the variety of amplitude profiles. 

There are interesting implications of the number of dimensions and their size for parameter estimation. Given an astrophysical signal, in the limit of high SNR $\rho$, the parameter likelihood is approximately given by $P(d \mid \bm{p}) \propto \exp(- \rho^2 |z|^2 / 2)$, where $z=(h(\bm{p_\ast}) \mid h(\bm p))$ is the complex match of $\bm{p}$ to the best-fit parameters $\bm{p_\ast}$~\cite{Roulet2019}. By virtue of Eq.~\eqref{eq:Euclidean}, this means that the likelihood is approximately an isotropic Gaussian in terms of the $c_\alpha$ coordinates, with a width $\sim 1/\rho$. The number of dimensions can therefore be interpreted as the number of independent parameters that can be measured, and the size of each dimension as the relative precision that can be obtained for a fixed SNR (with the caveat that we have restricted the frequency range; for example, information about the tidal deformability comes from frequencies higher than our \SI{512}{\hertz} cutoff). 

For example, for BNS (and effectively for light BBH) the banks have two dimensions, with a large first dimension well correlated with the chirp mass (Fig.~\ref{fig:calpha}). The two measurable parameters are the chirp mass, which indeed can be measured to much higher precision than for heavy systems, and a combination of the mass ratio and effective spin which can be measured with a lower precision. These are the leading contributions to the phase evolution as can be understood from the post-Newtonian expansion.

An important advantage of our geometric coordinates is that they are well suited for a two-step search that effectively achieves a smaller grid spacing at reduced computational cost. We realize this by refining the template grid on demand around all triggers that exceed an appropriately lowered SNR threshold~\cite{pipelinepaper, Gadre2018}. During the search, we first use a coarse grid, and refine every trigger using neighboring templates from a denser grid that has half the spacing along each dimension. The fact that the distance between $c_\alpha$ components translates directly to mismatch (Eq.~\eqref{eq:Euclidean}) makes this method straightforward to implement. 

\begin{figure}
    \centering
    \includegraphics[width=\linewidth]{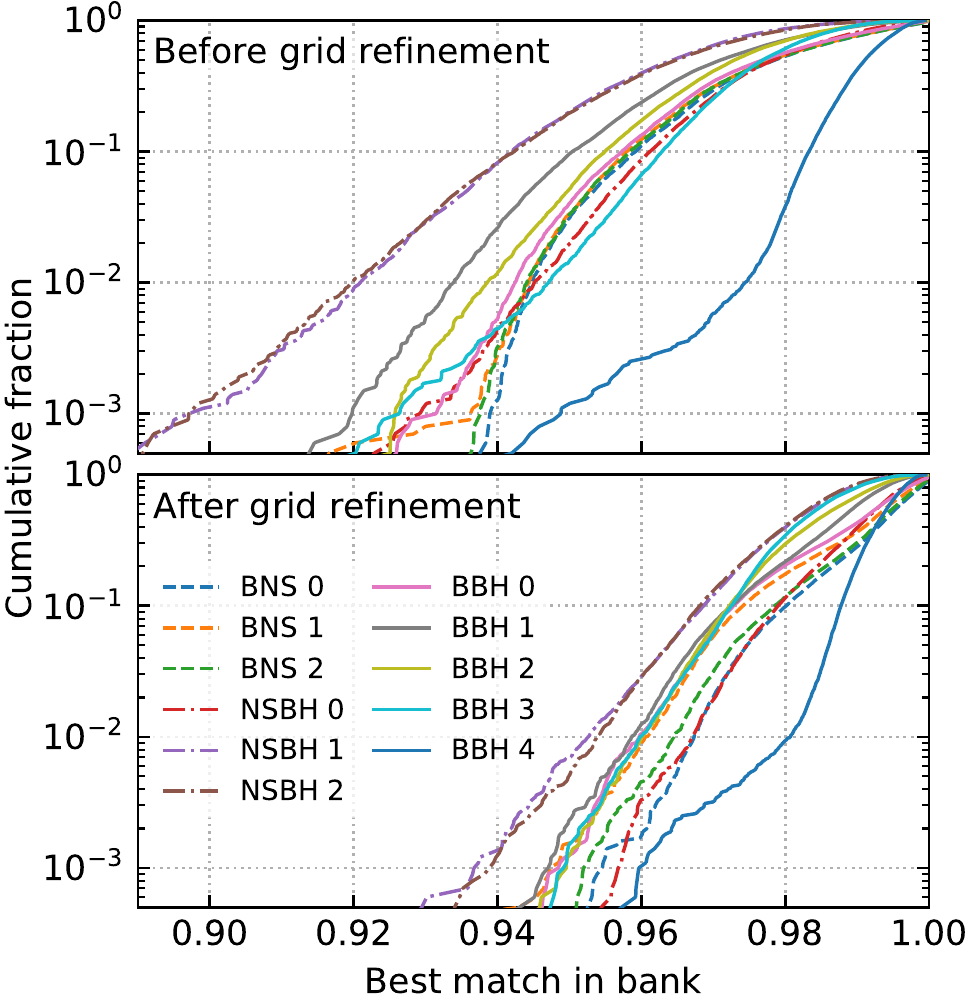}
    \caption{Effectualness of our template banks, tested on random waveforms drawn from a distribution uniform in individual masses and aligned spins. The vertical axis shows the fraction of the random trials that do not achieve a given match in the bank.}
    \label{fig:effectualness}
\end{figure}

To characterize the effectualness of the bank at recovering the target physical signals, we generate a set of \num{e4} random ``test waveforms'' within the parameter range of each bank, using the same approximant with which the input waveforms were generated. We choose the parameters from a distribution that is uniform in the component masses $m_1, m_2$ and aligned spins $\chi_1, \chi_2$.
In principle, we would have to match each test waveform against every waveform in the bank to look for the best match. To save computational effort, we select a candidate best-match based on the approximate metric Eq.~\eqref{eq:Euclidean} by extracting the phase of the test waveform $\psi^{(i)}(f)$, projecting it onto the linear space, $c_\alpha^{(i)} = \langle \psi^{(i)} - \overline \psi, \psi_\alpha \rangle$, and finding the closest grid point with respect to the Euclidean metric~\eqref{eq:Euclidean}. Since\textit{ a priori} we do not know which subbank best describes the test waveform, we pick the best candidate from each subbank and compute the match with all. The best match with our reduced set of candidates is a lower bound on the best match over all the waveforms in the bank.
Rather than using Eq.~\eqref{eq:match} directly, we compute the match by following the detection strategy described in \citet{pipelinepaper}: we account for the finite time resolution of the Fourier transform by downsampling the waveforms to \SI{512}{\hertz} and sinc-interpolating the matched-filter output twice.
We show the result of this test in Fig.~\ref{fig:effectualness}, in terms of the cumulative fraction of the matches with each bank before and after applying the grid refinement, which we use to assess the collection threshold on the coarse grid and the effectualness achieved for each bank, respectively. We find that depending on the bank 99\% of the templates have a match higher than 0.95 to 0.98.

\section{Conclusions}
\label{sec:conclusions}

We have developed a general and computationally efficient geometric placement algorithm to construct high-effectualness template banks for detecting gravitational waves from compact binary mergers. We have constructed a basis of functions that generate a linear space of phase profiles on which the mismatch metric is Euclidean. For the purpose of signal detection, we shift the focus away from physical parameters to the linear coefficients for the basis phase profiles.
We identified which components carry the largest amount of information about physical waveforms and what is the minimal set required to guarantee a desired match. 
The basis functions can be determined from a set of input waveforms whose size is small compared to that of the bank. The basis functions can be generated with any frequency-domain waveform model. The resolution of the bank can be decided independently after the basis functions have been found; in particular, it can be increased arbitrarily at negligible computational cost since no further evaluations of the physical waveform approximants need to be done.
Our algorithm guarantees that within each of the few subbanks that make up one template bank, all templates share the same amplitude profile, a property that is critical for the correction of the power-spectral-density drift in signal processing. 

We have applied our algorithm to the construction of a collection of eleven template banks that together cover the parameter space associated to stellar-mass compact binary mergers with aligned spins. We find the effectualness and total number of templates to be comparable to the ones obtained by other algorithms in the literature \citep{Canton2017, Roy2017b}; detailed comparisons are difficult due to the different parameter spaces targeted in various works. We note that our template bank includes rapidly spinning neutron stars, which to date have not been searched for in the gravitational wave data.
We implement a two-step search with a coarse grid that we refine around triggers at the time of search, a task for which our new formalism is ideally suited. This is an important step to reduce the number of templates while preserving a high effectualness.

Looking forward, an accurate and fast interpolation from physical parameters to the $c_\alpha$ component space would be extremely useful for rapid parameter estimation. First, because waveforms can be generated at negligible computational cost once the components are known. At least in cases where analytical waveform models are not valid, waveform generation dominates the computational cost of parameter estimation.
Moreover, the likelihood would look close to an isotropic Gaussian in terms of the $c_\alpha$ coordinates due to orthonormality, making them a suitable choice from the data analysis perspective.
\changed{Other natural extensions of the work presented here are to include the effects of precession, due to misalignment between the spins and the orbital angular momentum, and eccentricity. These are deferred for future work. The inclusion of eccentricity is currently limited by the availability of robust public waveform generation codes.}

The template bank described here is publicly available at \url{https://github.com/jroulet/template_bank}.

\section*{Acknowledgements}

This research has made use of data, software and/or web tools obtained from the Gravitational Wave Open Science Center (\url{https://www.gw-openscience.org}), a service of LIGO Laboratory, the LIGO Scientific Collaboration and the Virgo Collaboration. LIGO is funded by the U.S. National Science Foundation. Virgo is funded by the French Centre National de Recherche Scientifique (CNRS), the Italian Istituto Nazionale della Fisica Nucleare (INFN) and the Dutch Nikhef, with contributions by Polish and Hungarian institutes.

LD acknowledges the support from the Raymond and Beverly Sackler Foundation Fund.
TV acknowledges support by the Friends of the Institute for Advanced Study.
BZ acknowledges the support of The Peter Svennilson Membership fund.
MZ is supported by NSF grants AST-1409709,  PHY-1521097 and  PHY-1820775 the Canadian Institute for Advanced Research (CIFAR) program on Gravity and the Extreme Universe and the Simons Foundation Modern Inflationary Cosmology initiative.

\appendix
\section{Differences with the bank used in \citet{pipelinepaper}}
\label{app:tb_differences}

In this appendix we report the differences between the template bank presented in this work and the one that was actually used in the binary black hole search reported by \citet{pipelinepaper}.
\begin{itemize}
    \item The bank used in Ref.~\citep{pipelinepaper} was restricted to BBHs, with a narrower range for the aligned spins $\abs{\chi_{1,2}} < 0.85$ instead of $0.99$.
    \item The reference PSD used to build the bank was \texttt{aLIGO\_MID\_LOW} \citep{lalsuite} instead of an empirically measured one.
    \item As a consequence, the bank used a low-frequency cutoff at \SI{20}{\hertz} instead of \SI{24}{\hertz}. We found that with the empirical PSD, the relative contribution to squared SNR from frequencies below \SI{24}{\hertz} is $< 1\%$.
    \item The optimization of the subbank amplitude profiles with the $k$-means clustering algorithm was not done; instead the division into subbanks was computed in a way analogous to the ``stochastic placement'' approach to building a template bank described in \S\ref{sec:intro}, with a required amplitude match $(A_i \mid \overline A) > 0.95$. The reference amplitude of the subbank was given by the first waveform generated in the subbank. Using the $k$-means clustering improves the best match by $\sim 0.01$ on average.
\end{itemize}

\bibliographystyle{apsrev4-1-etal}
\bibliography{gw}

\end{document}